\documentclass[11pt]{article}
\usepackage{latexsym}
\usepackage{epsf}
\usepackage{times}
\usepackage{latexsym}
\usepackage{graphics}
\usepackage{graphicx}

\newtheorem{de}{Definition}

\newfont{\extra}{msbm10 scaled\magstep1}

\def\d(#1){\partial_{#1}}

\def\be{\begin{equation}}
\def\ee{\end{equation}}
\def\bea{\begin{eqnarray}}
\def\eea{\end{eqnarray}}

\def\p{\partial}
\def\a{\alpha}
\def\b{\beta}
\def\de{\delta}

\def\l{\lambda}

\def\bnabla{\mbox{\boldmath $\nabla$}}

\begin{document}

\begin{center}{ \LARGE \bf
Classical and quantum three-dimensional\\[2.ex]
   integrable  systems with axial symmetry}
\end{center}
\vskip0.25cm

\begin{center}
M. Gadella$^*$, J. Negro$^*$ and G. P. Pronko$^{\dagger}$

\vskip0.25cm

${}^{*}${\it  Departamento de  F\'{\i}sica Te\'orica, At\'omica y
\'Optica\\
Universidad de  Valladolid, E-47005, Valladolid,  Spain}

${}^{\dagger}${\it Institute of High Energy Physics, Protvino,\\
Moscow Region, 148280, Russia. \\
Institute of Nuclear Physics, NCSR, ``Demokritos'', Athens, Greece.
}

\vskip0.15cm

\end{center}

\vskip0.5cm \centerline{\today} \vskip0.5cm

\begin{abstract}
We study the most general form of a three dimensional classical
integrable system with axial symmetry and invariant under the axis
reflection. We assume that the three constants of motion are the
Hamiltonian, $H$, with the standard form of a kinetic part plus a
potential dependent on the position only, the $z$-component of the
angular momentum, $L$, and a Hamiltonian-like constant, $\widetilde
H$, for which the kinetic part is quadratic in the momenta. We find
the explicit form of these potentials compatible with complete
integrability. The classical equations of motion, written in terms
of two arbitrary potential functions, is separated in oblate
spheroidal coordinates. The quantization of such systems leads to a
set of two differential equations that can be presented in the form
of spheroidal wave equations.
\end{abstract}

\section{Introduction}

This is a paper on integrability of three dimensional systems, both
classical and quantum. Apart from being interesting by itself, this
study can be extremely interesting for a field, which is rather far
away from the present subject, the theory of the continuous media:
gas, fluid and plasma. Those are mechanical systems with an infinite
number of degrees of freedom and certainly much more difficult than
the mechanics of a particle in three dimensions. However, in the
theory of continuous media, we can pose a problem of the existence
of solitons: the steady solution with time-independent field with
density $\rho(x)$ and velocities ${\bf v}(x)$. It could exist in the
non dissipative case only, as we neglect the viscosity. In this
latter case, the particles which constitute the media should move in
a self consistent potential along its trajectories, which should be
closed. Although this potential is the result of interparticle
interaction, each single particle moves in an effective potential
that provides the closed trajectory. From the analysis of the
equations of motion in continuous media (\cite{P}), it follows that
the simplest shape of solution is a toroid. Thus, we have to look
for a three dimensional single particle integrable system including
a toroidal shape. One of the goals of the present paper is to make a
first step in this direction, as we describe all completely
integrable systems with axial symmetry. The next step would be to
select among all the completely integrable systems those sharing
some specific properties.

\section{The axial coordinates}\label{integrable}

Let us consider a three dimensional classical system with the
canonical coordinates $(p_i,x_i)$, $i=1,2,3$, where the commutators
are defined by the usual Poisson brackets $\{A,B\}=\frac{\partial A}
{\partial p_k}\frac{\partial B}{\partial x_k}-\frac{\partial A}
{\partial x_k}\frac{\partial B}{\partial p_k}$, the sum being
understood over repeated indexes. The evolution of the system is
described by the Hamiltonian

\be\label{h} H= \frac{{\bf p}^2}{2m} + U({\bf x}),\qquad \qquad {\bf
x}=(x_1,x_2,x_3),\  {\bf p}=(p_1,p_2,p_3)\,,
\ee
where $U({\bf x})$
is a potential term. Let us assume the existence of two additional
integrals of motion for this system, all of them in involution. The
first one will be chosen as the angular momentum along the direction
fixed by the unit vector ${\bf n}=(n_1,n_2,n_3)$:

\be\label{l} L = {\bf n}\cdot\left({\bf x}\times {\bf p}\right)
\ee
As for the second one we require to be quadratic in momentum and
commuting with both, $H$ and $L$. Then, it is straightforward to
show that, provided we include also the reflection with respect to
the ${\bf n}$-axis as a discrete symmetry of our system, it must
take the general expression:

\be\label{hbis} \widetilde H=\frac{1}{2m}\,p_i\,g^{ik}({\bf x})\,p_k
+ \Phi({\bf x})
\ee
where the quadratic term has the `metric'

\be\label{g} g^{ik}({\bf x})= \de^{ik} ({{\bf x}\cdot{\bf n}})^2 -
x_{\bf n} (x_i n_k + x_k n_i) + ({\bf x}^2- a^2)n_i n_k\,,
\ee
being
$a^2$ a real constant that we take positive. The quadratic term in
the momenta can be written as

\be\label{htris} \frac1{2m}\ (L_{\perp}^2 - a^2 {p_{\bf n}}^2),
\ee
where $L_{\perp}$ is the perpendicular component of ${\bf L}$ to the
${\bf n}$-axis, and $p_{\bf n}= {\bf p}\cdot {\bf n}$. Here we must
point out that the operators (\ref{l}) and (\ref{htris}) determine
the prolate-oblate spheroidal coordinates that separate the
Laplacian operator \cite{miller}.

The commutation of $L$ with $H$ and $\widetilde H$ restrict the form
of their potential terms

\be\label{uf} U({\bf x}) = U({\bf x}^2,({{\bf x}\cdot{\bf
n}})^2),\qquad \Phi({\bf x}) = \Phi({\bf x}^2,({{\bf x}\cdot{\bf
n}})^2)
\ee
while the commutation of $H$ and $\widetilde H$ leads to
the equations

\be\label{potenciales} \d(i)\Phi({\bf x}) = g^{ik}({\bf x})\, \d(k)
U({\bf x})
\ee
In order to deal with (\ref{potenciales}) we
diagonalize the matrix $g^{ik}({\bf x})$. Its eigenvalues $\l({\bf
x})$ and eigenvectors $A({\bf x})$ are obtained from the matrix
equation

\be\label{matrix} \left(g^{ik}({\bf x})- \l({\bf x}) \de^{ik}\right)
A_k({\bf x})=0
\ee
We get the solutions

\be\label{lambdas} \l_{\pm}({\bf x}) = \frac{{\bf
x}^2-a^2}{2}\pm\sqrt{\left(\frac{{\bf x}^2-a^2}{2}\right)^2+ a^2\,
({{\bf x}\cdot{\bf n}})^2},\qquad \l_0({\bf x})= ({{\bf x}\cdot{\bf
n}})^2
\ee
The corresponding eigenvectors can be expressed in the
following way:

\be A^+_i({\bf x})=\d(i)\l_-({\bf x}),\quad A^-_i({\bf
x})=\d(i)\l_+({\bf x}),\quad A^0_i({\bf x})= \frac{{\bf n}\times{\bf
x}}{{\bf x}^2-({{\bf x}\cdot{\bf n}})^2}= \d(i)\varphi({\bf x})
\label{2.9}
\ee
where $\varphi$ is the azimutal angle around the
${\bf n}$-axis. Let us write here some useful identities of these
eigenvalue functions for future calculations,

\be
\begin{array}{l}
\displaystyle (\d(i)\l_+)^2 =
\frac{4\l_+}{\l_+-\l_-}\left(\l_++a^2\right)
\\[2.5ex]
\displaystyle (\d(i)\l_-)^2 =
-\frac{4\l_-}{\l_+-\l_-}\left(\l_-+a^2\right)
\\[2.5ex]
\displaystyle (\d(i)\l_0)^2 =-\frac{4}{a^2}\,\l_+\l_-
\end{array}
\ee

Since the eigenvectors (\ref{2.9}) are orthogonal, it is natural to
adopt as new orthogonal coordinates the set $\{\l_+({\bf
x}),\l_-({\bf x}),\varphi({\bf x})\}$, which are essentially the
{\it oblate spheroidal} coordinates \cite{miller}. We turn to eq.\
(\ref{potenciales}) for the potentials, now expressed in the new
coordinate system. Taking into account that due to the geometric
symmetry, $L$ (\ref {l}),  $U$ and $\Phi$ do not depend on
$\varphi$, (\ref {potenciales}) becomes

\be \d(i)\l_+\,\d(+)\Phi + \d(i)\l_-\,\d(-)\Phi=
g^{ik}\left(\d(k)\l_+\,\d(+)U + \d(k)\l_-\,\d(-)U\right)
\ee
where
$\d(\pm)$ stand for $\frac{\partial}{\partial \l_{\pm}}$. Now, as
$g^{ik}$ is diagonal in the new coordinate basis $\{
\mathbf{\bnabla}\l_-,\mathbf{ \bnabla}\l_+,\mathbf{
\bnabla}\varphi\}$, this equation decouples in

\be \d(+) \Phi = \l_- \d(+)U,\qquad \d(-) \Phi = \l_+ \d(-)U \ee or
\be\label{philambda} \d(+) (\Phi - \l_- U)=0,\qquad \d(-) (\Phi -
\l_+ U)=0
\ee
This means that

\be\label{fg} \Phi - \l_+ U = -f(\l_+),\qquad   \Phi - \l_- U =
-g(\l_-)
\ee
where $f(\l_+)$ and $g(\l_-)$ are arbitrary functions.
The minus sign in the front of (\ref{fg}) intends that the next
equations, derived from (\ref{fg}), can be written in a symmetric
form with respect to the variables $\lambda_+$ and $\lambda_-$.
These equations give an expression for the potentials as follows:

\be\label{uphi} U=\frac{f(\l_+)-g(\lambda_-)}{\l_+-\l_-},\qquad
\Phi=\frac{\l_-}{\l_+-\l_-}\,f(\l_+) -
\frac{\l_+}{\l_+-\l_-}\,g(\l_-)\,. \ee These are the most general
expressions for the potentials $U({\bf x})$ and $\Phi({\bf x})$
compatible with $\{H,\widetilde H\}=0$.

\section{Separation of variables}

\subsection{The momentum as a gradient}

Now we will recall here a property for a classical system in three
dimensions having three integrals of motion (including the
Hamiltonian) $h_i$, $i=1,2,3$, in involution, thus being integrable:

\bea
&&h_i({\bf p},{\bf x})=e_i,\qquad \qquad  i=1,2,3\label{haches}\\[2.ex]
&&\{h_i,h_j\}\equiv \frac{\partial h_i}{\partial p_k}\frac{\partial
h_j}{\partial x_k}-\frac{\partial h_i}{\partial x_k}\frac{\partial
h_j}{\partial p_k}=0,\qquad \forall i,j \label{3.17}\eea If we
assume that the determinant of the matrix $\left(\frac{\partial
h_i}{\partial p_k}\right)$ is nonvanishing, the inverse function
theorem gives from (\ref{haches}) the expression of the momenta as
functions of coordinates, at least locally:

   \be p_i=f_i({\bf x},{\bf
e}) \ee so that \be h_i({\bf f}({\bf x},{\bf e}),{\bf x}) = e_i \ee
Taking partial derivatives of this equation, we obtain \bea
&&\frac{\partial h_i}{\partial p_k}\frac{\partial f_k}{\partial
x_l}+ \frac{\partial h_i}{\partial x_l}=0 \quad \Longrightarrow
\label{h1}\\[2.ex]
&& \frac{\partial h_i}{\partial p_k}\frac{\partial f_k}{\partial
x_l} \frac{\partial h_m}{\partial p_l}= -\frac{\partial
h_i}{\partial x_l}\frac{\partial h_m}{\partial p_l}= -\frac{\partial
h_m}{\partial x_l}\frac{\partial h_i}{\partial p_l} \quad
\Longrightarrow
\label{3.21}\\[2.ex]
&& \frac{\partial h_i}{\partial p_k}\frac{\partial f_k}{\partial
x_l} \frac{\partial h_m}{\partial p_l}+ \frac{\partial h_i}{\partial
p_k}\frac{\partial h_m}{\partial x_k}=0 \eea Note that the second
identity in (\ref{3.21}) comes after condition (\ref{3.17}).
According to our hypothesis, the determinant of the matrix
$\left(\frac{\partial h_i}{\partial p_k}\right)$ vanishes, so that
the last relation can be simplified:

\be\label{h2} \frac{\partial f_k}{\partial x_l} \frac{\partial
h_m}{\partial p_l}+ \frac{\partial h_m}{\partial x_k}=0 \ee From
(\ref{h1}) and (\ref{h2}) we have \be \frac{\partial h_m}{\partial
p_l} \left(\frac{\partial f_k}{\partial x_l}- \frac{\partial
f_l}{\partial x_k}\right)=0 \ee hence \be \frac{\partial
f_k}{\partial x_l}= \frac{\partial f_l}{\partial x_k} \ee This means
that the vector field, in the variable $\bf x$, ${\bf f}({\bf
x},{\bf e})$ has vanishing rotational, i.e.,

\begin{equation}\label{3.26}
     {\rm\bf rot}_{\bf x}\, {\bf f}({\bf x},{\bf e})={\bf 0}\,.
\end{equation}
Therefore, if ${\bf f}({\bf x},{\bf e})$ is sufficiently regular,
there exists a function $F({\bf x}, {\bf e})$, locally defined, such
that

\begin{equation}\label{3.27}
f_l=\frac{\partial F({\bf x}, {\bf e})}{\partial x_l} \hskip1cm{\rm
i.e.}\hskip1cm {\rm\bf grad}_{\bf x}\, F({\bf x}, {\bf e}) = {\bf
f}({\bf x},{\bf e})\,.
\end{equation}

In conclusion, we have shown that if there are three integrals of
motion (\ref{haches}) in involution, the momentum will be the
gradient of the function $F$: \be\label{F} p_l=\frac{\partial F({\bf
x}, {\bf e})}{\partial x_l} \ee These could be considered as a first
class constraints. The function $F$ is the characteristic function
in the Hamilton-Jacobi approach. Note that this is a general
property, valid for any dimension $n$.

\subsection{The separation of $F$}

We can apply the above results to our system by making the
identification $h_1=H$, $h_2=L$, $h_3=\widetilde H$. The next step
is to show the separability  of the function $F$ in the variables
$\{\l_\pm, \varphi\}$ introduced in the previous section. Then, if
we apply the chain rule to (\ref{3.27}) and make use of (\ref{2.9}),
we have

\be {\bf p} = \mathbf{\bnabla}\l_+\, \d(+)F(\l_+,\l_-,\varphi)+
\mathbf{\bnabla}\l_-\, \d(-)F(\l_+,\l_-,\varphi)+ \frac{{\bf
n}\times{\bf x}}{{\bf x}^2-({{\bf x}\cdot{\bf n}})^2}\,
\d(\varphi)F(\l_+,\l_-,\varphi)\,. \ee
Since the vector fields in
(\ref{2.9}) are mutually orthogonal, multiplying both sides of the
above expression by ${\bf n}\times{\bf x}$, we obtain

\be\label{fphi} \left({\bf n}\times{\bf x}\right)\cdot {\bf p} =
{\bf n}\cdot {\bf L} = \ell=\d(\varphi)F(\l_+,\l_-,\varphi)\,, \ee
where $\ell$ is the value of the integral of motion $L$
corresponding to the angular momentum around ${\bf n}$.  We conclude
that

\be\label{pes} {\bf p} = \mathbf{\bnabla}\l_+\,
\d(+)F(\l_+,\l_-,\varphi)+ \mathbf{\bnabla}\l_-\,
\d(-)F(\l_+,\l_-,\varphi)+ \frac{{\bf n}\times{\bf x}}{{\bf
x}^2-({{\bf x}\cdot{\bf n}})^2}\, \ell \ee and therefore taking the
square modulus in (\ref{pes}), we have that

\be\label{p2} {\bf p}^2 = (\mathbf{\bnabla}\l_+)^2 (\d(+)F)^2+
(\mathbf{\bnabla}\l_-)^2 (\d(-)F)^2+ \frac{\ell^2}{{\bf x}^2-({{\bf
x}\cdot{\bf n}})^2}= 2m\left(E-U\right) \ee
where $E$ is the value
of the constant of motion $H$. We recall that the expression
(\ref{pes}) for the momentum vector field has been obtained in the
basis that diagonalizes the metric matrix of components
$\{g^{ik}\}$. In this basis and from (\ref{pes}), we
straightforwardly compute: \be\label{pgp} p_ig^{ik}p_k =
\l_-(\bnabla \l_+)^2 (\d(+)F)^2+ \l_+(\bnabla\l_-)^2 (\d(-)F)^2+
\frac{({{\bf x}\cdot{\bf n}})^2\,\ell^2}{{\bf x}^2-({{\bf
x}\cdot{\bf n}})^2}=2m(\widetilde E - \Phi) \ee where $\widetilde E$
is the value of $\widetilde H$.

Next, we multiply equation (\ref{p2}) by $\l_+$, (\ref{pgp}) by $-1$
and sum. Next we make the same procedure with $\lambda_-$ instead.
Finally, we get the following two expressions:

\be\label{multiply} \left\{
\begin{array}{l}
\displaystyle(\mathbf{\bnabla}\l_+)^2(\l_+-\l_-) (\d(+)F)^2 =
2m\left(\l_+(E-U)-(\widetilde E-\Phi)\right) -\frac{\ell^2}{{\bf
x}^2-({{\bf x}\cdot{\bf n}})^2}(\l_+-({{\bf x}
\cdot{\bf n}})^2)\\[2.ex]
-\displaystyle(\mathbf{\bnabla}\l_-)^2(\l_+-\l_-) (\d(-)F)^2 =
2m\left(\l_-(E-U)-(\widetilde E-\Phi)\right) -\frac{\ell^2}{{\bf
x}^2-({{\bf x}\cdot{\bf n}})^2}(\l_--({{\bf x}\cdot{\bf n}})^2)
\end{array}
\right. \ee
Taking into account the definition (\ref{lambdas}) of
$\l_\pm$, the next formulas are easily obtained:

\be\label{dos}
\begin{array}{l}
{\bf x}^2 - ({{\bf x}\cdot{\bf n}})^2 =\l_++ \l_-
+a^2+\frac{\l_+\l_-}{a^2} =
\frac{1}{a^2} \, (\l_++a^2)(\l_-+a^2)\\[2.ex]
\l_+-({{\bf x}\cdot{\bf n}})^2=\l_++\frac{\l_+\l_-}{a^2}=\frac{\l_+}
{a^2}\,(\l_-+a^2)\\[2.ex]
\l_--({{\bf x}\cdot{\bf n}})^2=\l_-+\frac{\l_+\l_-}{a^2}=\frac{\l_-}
{a^2}\,(\l_++a^2)\\[2.ex]
(\l_+-\l_-)(\bnabla \l_+)^2= 4\l_+(\l_++a^2)\\[2.ex]
(\l_+-\l_-)(\bnabla \l_-)^2= -4\l_-(\l_-+a^2)\,.
\end{array}
\ee Then, we carry (\ref{dos}) into (\ref{multiply}) to get

\be
\begin{array}{l}
\displaystyle 4\l_+(\l_++a^2) (\d(+)F)^2 =
2m\left(\l_+(E-U)-(\widetilde E-\Phi)\right)
-\frac{ \l_+\, \ell^2}{\l_+ +a^2}\\[2.ex]
\displaystyle 4\l_-(\l_-+a^2) (\d(-)F)^2 =
2m\left(\l_-(E-U)-(\widetilde E-\Phi)\right) -\frac{ \l_-\,
\ell^2}{\l_- +a^2}\,.
\end{array}\label{3.36}
\ee We finally have \be\label{fs}
\begin{array}{l}
\displaystyle (\d(+)F)^2 =
\frac{1}{4\l_+(\l_++a^2)}\left\{2m\left(\l_+(E-U)-(\widetilde E-
\Phi)\right)
-\frac{ \l_+\, \ell^2}{\l_+ +a^2}\right\}\\[2.ex]
\displaystyle (\d(-)F)^2 =
\frac{1}{4\l_-(\l_-+a^2)}\left\{2m\left(\l_-(E-U)-(\widetilde E-
\Phi)\right) -\frac{ \l_-\, \ell^2}{\l_- +a^2}\right\}
\end{array}
\ee Now, let us observe that in the right hand side of (\ref{fs})
the combination $\l_+(E-U)-(\widetilde E-\Phi)$ cannot depend on
$\l_-$. At the same time,  $\l_-(E-U)-(\widetilde E-\Phi)$ will not
depend on $\l_+$. Note that a similar behavior have arosen in
(\ref{philambda}). Hence, since $\d(+)F$ and $\d(-)F$ depends only
on $\l_+$ and $\l_-$, respectively, while the dependence on
$\varphi$ was given by (\ref{fphi}), we conclude that we have
managed to obtain the separation of $F$ in the variables $\l_\pm,
\varphi$, i.e., \be\label{fsep} F(\l_+,\l_-,\varphi)=A(\l_+) +
B(\l_-) + \ell\, \varphi \ee Later, we shall find an explicit form
for $F$ in terms of a different set of variables.

\subsection{The integration of the equations of motion}

Our next objective is the integration of the following equations of
motion: \be p_i= \d(i) F({\bf x})\,. \label{3.39} \ee In other
words, our goal is finding the explicit dependence of the variables
$\lambda_\pm$ and $\varphi$ with time, i.e., the functions
$\lambda_\pm=\lambda_\pm(t)$ and $\varphi=\varphi(t)$. From
(\ref{fs}) and (\ref{fsep}), it is clear that

\be
\begin{array}{l}
\left(\d(+)A(\l_+)\right)^2= P({\bf x},\l_+)\\[2.ex]
\left(\d(-)B(\l_-)\right)^2= P({\bf x},\l_-)\,,
\end{array}
\ee where we are using the shorthand notation \be P({\bf x},\l_\pm)
= \frac{1}{4\l_\pm(\l_\pm+a^2)}\left\{2m\left(\l_\pm(E-U({\bf
x}))-(\widetilde E-\Phi({\bf x}))\right) -\frac{\ell^2
\l_\pm}{\l_\pm +a^2}\right\}\,. \ee The chain rule along ${\bf p} =
m\,\dot{\bf x}$ and (\ref{3.39}) give us

\be
\begin{array}{l}
\displaystyle \dot\l_+=\d(i)\l_+ \, \dot x_i= \frac1m\, \d(i)\l_+ \,
p_i= \frac1m\, \d(i)\l_+ \,
\partial_iF\\[2.ex]
\displaystyle \dot\l_-=\d(i)\l_- \, \dot x_i= \frac1m\, \d(i)\l_- \,
p_i= \frac1m\,
\d(i)\l_- \, \partial_iF\\[2.ex]
\displaystyle \dot \varphi =\d(i) \varphi\, \dot x_i= \frac1{{\bf
x}^2 -({\bf x}\cdot{\bf n})^2}\, \frac{\ell}{m}=\frac
1m\partial_i\varphi\,\partial_iF\,,
\end{array} \label{3.42}
\ee where the dot denotes time derivative, as usual. Therefore,

\be \left\{
\begin{array}{l}
\displaystyle \dot \l_+ = \frac1m\, \d(i)\l_+ \d(i)F=\frac1m\,
(\d(i)\l_+)^2\d(+)A(\l_+)=\frac1m\,\frac{4\l_+}{\l_+-\l_-}(\l_++a^2)\d
(+)A(\l_+)\\[2.ex]
\displaystyle \dot \l_-
=-\frac1m\,\frac{4\l_-}{\l_+-\l_-}(\l_-+a^2)\d(-)B(\l_-)
\\[2.ex]
\displaystyle \dot \varphi= \frac{a^2 \ell}{m}\,\frac{1}{(\l_+
+a^2)(\l_-+a^2)} =\frac{a^2 \ell}{m}\, \frac1{\l_+-\l_-}
\left(\frac{1}{\l_-+a^2}-\frac{1}{\l_++a^2}\right)\,.
\end{array}
\right. \label{3.43} \ee
 From the first two equations in
(\ref{3.43}),

\be\label{ldots} \left\{
\begin{array}{l}
\displaystyle \frac{\dot \l_+}{\l_+\d(+)A(\l_+)}=
\frac1m\, \frac{4(\l_++a^2)}{\l_+-\l_-}\\[2.5ex]
\displaystyle \frac{\dot \l_-}{\l_-\d(-)B(\l_-)}= -\frac1m\,
\frac{4(\l_-+a^2)}{\l_+-\l_-}
\end{array}
\right.\,, \ee we can get three expressions separated in $\l_+$,
$\l_-$ and $\varphi$:

\be\label{derivatives} \left\{
\begin{array}{l}
\displaystyle \frac{\dot \l_+}{\l_+\d(+)A(\l_+)}+ \frac{\dot
\l_-}{\l_-\d(-)B(\l_-)}=
\frac4m \\[2.5ex]
\displaystyle \frac{\dot \l_+}{\l_+(\l_++a^2)\d(+)A(\l_+)}+
\frac{\dot \l_-}{\l_-(\l_-+a^2)\d(-)B(\l_-)}=
0\\[2.5ex]
\displaystyle -\ell a^2\left(\frac{\dot
\l_+}{4\l_+(\l_++a^2)^2\d(+)A(\l_+)}+ \frac{\dot
\l_-}{4\l_-(\l_-+a^2)^2\d(-)B(\l_-)}\right)= \dot\varphi
\end{array}
\right. \ee If the potentials (\ref{uphi}) are known, e.g., the
functions $f(\l_+)$ and $g(\l_-)$ are chosen, then the functions
$\d(+)A(\l_+)$ and $\d(-)B(\l_-)$ will be also explicitly known.
Thus, integrating over time (\ref{derivatives}) we obtain the
following crude expressions:

\be \left\{
\begin{array}{l}
\displaystyle \int_0^{\l_+}{\rm d}z\,\frac{1}{z\,\d(z)A(z)}+
\int_0^{\l_-}{\rm d}z\,\frac{1}{z\,\d(z)B(z)}=
\frac4m\,t + c_1 \\[3.ex]
\displaystyle \int_0^{\l_+}{\rm d}z\,\frac{1}{z(z+a^2)\d(z)A(z)}+
\int_0^{\l_-}{\rm d}z\,\frac{1}{z(z+a^2)\d(z)B(z)}=
c_2\\[2.5ex]
\displaystyle -\ell a^2\left(\int_0^{\l_+}{\rm d}z\,
\frac{1}{4z(z+a^2)^2\d(z)A(z)}+ \int_0^{\l_-}{\rm d}z\,
\frac{1}{4z(z+a^2)^2\d(z)B(z)}\right)= \varphi + c_3
\end{array}
\right. \ee These relations constitute an implicit form of the
integration of the equations of motion. To obtain the coordinates
$\l_+$, $\l_-$ and $\varphi$ as explicit functions of time we should
invert such relations, a task which is not often simple and that
will depend on each particular choice of the functions
$f(\lambda_+)$ and $g(\lambda_-)$.

\subsection{The function $F$ in terms of oblate spheroidal coordinates}

 From the definition (\ref{lambdas}) $\lambda_+$ is always positive
meanwhile the values of $\lambda_-$ lie on the interval
$[-a^2,0]$\,, depending on the scalar product ${\bf x}\cdot{\bf n}$.
Thus, we suggest the following change of coordinates:

\be \l_+ = a^2 \sinh^2\a,\qquad \l_- = -a^2 \sin^2 \b
\label{3.47}\,. \ee
Therefore we have \be x=a\cosh \alpha\sin \beta
\cos\varphi,\quad y = a \cosh\alpha\sin \beta\sin\varphi, \quad z =
a \sinh \alpha\cos\beta \ee where $\{\alpha,\beta,\varphi\}$ are the
usual oblate spherical coordinates \cite{stratton}-\cite{miller}.
Here, we shall see how the dependence of $F$ in terms of $\alpha$,
$\beta$ and $\varphi$ gives a new insight into the above discussion.
If we take time derivative in (\ref{3.47}), we obtain:

\begin{equation}\label{3.48}
     \dot\lambda_+=(2a^2\sinh\alpha \cosh\alpha)\,\dot\alpha
     \hskip1cm;\hskip1cm
     \dot\lambda_-=-(2a^2\sin\beta\cos\beta)\,\dot\beta\,.
\end{equation}
 From (\ref{3.47}), one readily obtains

\begin{eqnarray}
   4\lambda_+(\lambda_++a^2)\,\left(\frac{\partial
F}{\partial\lambda_+}
    \right)^2 &=& \left( \frac{\partial F}{\partial\alpha}\right)^2
\label{3.49}
    \\[2ex]
   4\lambda_-(\lambda_-+a^2)\,\left(\frac{\partial
F}{\partial\lambda_-}
    \right)^2 &=& - \left( \frac{\partial
    F}{\partial\beta}\right)^2\,. \label{3.50}
\end{eqnarray}
Now, we compare the expressions given in (\ref{3.42}) and
(\ref{3.47}) for $\dot\lambda_+$ and use (\ref{3.49}) to conclude
that

\begin{equation}\label{3.51}
     \dot\alpha=\frac 1m\;
     \frac{1}{\sinh^2\alpha+\sin^2\beta}\;\frac{\partial F}{\partial
     \alpha}\,.
\end{equation}
An analogous manipulation shows that

\begin{equation}\label{3.52}
     \dot\beta=\frac 1m\;\frac{1}{\sinh^2\alpha+\sin^2\beta}\;
     \frac{\partial F}{\partial \beta}\,.
\end{equation}
Using the last formula in (\ref{3.43}) and (\ref{3.47}), we find the
expression for the time derivative of $\varphi$ as

\begin{equation}\label{3.53}
     \dot\varphi=\frac{\ell}{ma^2}\;
     \frac{1}{\sinh^2\alpha+\sin^2\beta}\;\left(
     \frac{1}{\cos^2\beta}-\frac{1}{\cosh\alpha}\right)\,.
\end{equation}
Since the functions $f(\cdot)$ and $g(\cdot)$ depend respectively of
$\lambda_+$ and $\lambda_-$ only, they can be written as functions
of $\alpha$ and $\beta$, respectively. As $f(\cdot)$ and $g(\cdot)$
are, in principle, arbitrary, we could denote these functions as
$f(\alpha)$ and $g(\beta)$ respectively. Thus, (\ref{fg}) can be
written as (except for an irrelevant change on the sign):

\begin{equation}\label{3.54}
     \Phi-a^2(\sinh^2\alpha)\, U=f(\alpha),\hskip1cm
     \Phi+a^2(\sin^2\beta)\, U=g(\beta)\,.
\end{equation}
Then, (\ref{3.36}) with (\ref{3.49}-\ref{3.50}) and (\ref{3.54})
give

\begin{eqnarray}
   \left(\frac{\partial F}{\partial \alpha}\right)^2 &=&
   2m\left[a^2\sinh^2\alpha\,E-\widetilde E +f(\alpha) \right]-\ell^2
\;\frac{\sinh^2\alpha}{\cosh^2\alpha}
   \label{3.55}\\[2ex]
   \left(\frac{\partial F}{\partial \beta}\right)^2 &=& 2m \left[a^2
\sin^2\beta\,E+\widetilde E -g(\beta)
   \right]-\ell^2\;\frac{\sin^2\beta}{\cos^2\beta}\,,
   \label{3.56}
\end{eqnarray}
so that, we finally get the following expressions for the
derivatives of $\alpha$ and $\beta$:

\begin{eqnarray}
\dot\alpha &=& \frac {1}{a^2m} \;
\frac{1}{\sinh^2\alpha+\sin^2\beta} \; \left[
2m\left[a^2\sinh^2\alpha\,E-\widetilde E +f(\alpha) \right]-\ell^2\;
\frac{\sinh^2\alpha}{\cosh^2\alpha}\right]^{1/2}\nonumber\\[2ex]
&=&\frac {1}{a^2m} \; \frac{1}{\sinh^2\alpha+\sin^2\beta} \; \Delta_
\alpha^{1/2}\label{3.57}\\
[2ex]
   \dot\beta &=& \frac {1}{a^2m}\; \frac{1}{\sinh^2\alpha+\sin^2
\beta} \;
   \left[2m \left[a^2\sin^2\beta\,E+\widetilde E -g(\beta)
   \right]-\ell^2\;\frac{\sin^2\beta}{\cos^2\beta}
   \right]^{1/2} \nonumber\\[2ex]
   &=& \frac {1}{a^2m} \; \frac{1}{\sinh^2\alpha+\sin^2\beta} \;
\Delta_\beta^{1/2}\,, \label{3.58}
\end{eqnarray}
where obviously,

\begin{eqnarray}
   \Delta_\alpha &=& 2m\left[a^2\sinh^2\alpha\,E-\widetilde E
+f(\alpha) \right]-\ell^2\;
\frac{\sinh^2\alpha}{\cosh^2\alpha} \label{DA} \\[2ex]
   \Delta_\beta &=& 2m \left[a^2\sin^2\beta\,E+\widetilde E -g(\beta)
   \right]-\ell^2\;\frac{\sin^2\beta}{\cos^2\beta}\,.\label{DB}
\end{eqnarray}
Clearly, after (\ref{3.57}) and (\ref{3.58}), we have

\begin{equation}\label{3.61}
     \frac{\dot\alpha}{\Delta_\alpha^{1/2}}-\frac{\dot\beta}{\Delta_
\beta^{1/2}}=0
     \Longrightarrow \int_0^\alpha
     \frac{d\alpha}{\Delta_\alpha^{1/2}}-\int_0^\beta
     \frac{d\beta}{\Delta_\beta^{1/2}}=C_2\,,
\end{equation}
where $C_2$ is a constant with respect to time (obviously, it
depends on $\alpha$ and $\beta$). Another constant of motion can be
obtained as follows: First, we write (\ref{3.53}) as

\begin{equation}\label{3.62}
     \dot\varphi=\frac{\ell}{ma^2}\;
     \frac{1}{\sinh^2\alpha+\sin^2\beta}\;\left(
     \frac{\cos^2\beta+\sin^2\beta}{\cos^2\beta}-\frac{\cosh^2\alpha-
\sinh^2\alpha}{\cosh\alpha}\right)
\end{equation}
Then, using (\ref{3.57}) and (\ref{3.58}) in (\ref{3.62}), we have

\begin{equation}\label{3.63}
   \dot\varphi=  \frac{\ell}{a^2}\left(\frac{\dot\beta}{\Delta_\beta^
{1/2}}\;\frac{\sin^2\beta}{\cos^2\beta}
     +\frac{\dot\alpha}{\Delta_\alpha^{1/2}}\;\frac{\sinh^2\alpha}
{\cosh^2\alpha}
     \right)\,.
\end{equation}
 From (\ref{3.63}), we obtain

\begin{equation}\label{3.64}
     \frac{d}{dt}\left[ \varphi-\frac{\ell}{a^2} \left( \int
     \frac{d\alpha}{\Delta_\alpha^{1/2}}
     \;\frac{\sinh^2\alpha}{\cosh^2\alpha}+\int \frac{d\beta}{\Delta_
\beta^{1/2}} \;\frac{\sin^2\beta}{\cos^2\beta} \right)
     \right]=0\,,
\end{equation}
which shows that

\begin{equation}\label{3.65}
\varphi-\frac{\ell}{a^2} \left( \int
     \frac{d\alpha}{\Delta_\alpha^{1/2}}
     \;\frac{\sinh^2\alpha}{\cosh^2\alpha}+\int \frac{d\beta}{\Delta_
\beta^{1/2}} \;\frac{\sin^2\beta}{\cos^2\beta} \right)=C_3
\end{equation}
is a constant of motion.

Furthermore, from (\ref{3.57}) and (\ref{3.58}) we can obtain a
third result as we show in the sequel

\begin{equation}\label{3.66}
     \frac{\dot\alpha}{\Delta_\alpha^{1/2}}\;\sinh^2\alpha+
     \frac{\dot\beta}{\Delta_\beta^{1/2}}\;\sin^2\beta=
     \frac {1}{a^2m}\;\frac{\sinh^2\alpha}{\sinh^2\alpha+\sin^2\beta}+
     \frac {1}{a^2m}\;\frac{\sin^2\beta}{\sinh^2\alpha+\sin^2\beta}=
\frac 1m\,,
\end{equation}
which obviously yields after integration

\begin{equation}\label{3.67}
     \int \frac{d\alpha}{\Delta_\alpha^{1/2}}\;\sinh^2\alpha + \int
\frac{d\beta}{\Delta_\beta^{1/2}}\;\sin^2\beta= \frac {1}{a^2m}\;
t\,.
\end{equation}
The function $F$ can now be written in terms of the variables
$(\alpha,\beta,\varphi)$. Since $\lambda_+$ and $\lambda_-$ are
functions of  $\alpha$ and $\beta$ alone, respectively, formula
(\ref{fsep}) can be written as
\begin{equation}\label{3.68}
F(\alpha,\beta,\varphi)=A(\alpha)+B(\beta)+\ell\varphi
\end{equation}
where

\begin{equation}\label{3.69}
     A(\alpha)= \int \frac{\partial F}{\partial\alpha}\,d\alpha\,,
     \hskip1cm B(\beta)=\int \frac{\partial
     F}{\partial\beta}\,d\beta
\end{equation}
This gives the final expression for $F(\alpha,\beta,\varphi)$ as

\begin{equation}\label{3.70}
F(\alpha,\beta,\varphi)= \int \Delta_\alpha^{1/2} \,d\alpha
+\int\Delta_\beta^{1/2}\,d\beta+\ell\varphi\,.
\end{equation}
Time invariants $C_2$ and $C_3$ can be written in terms of certain
partial derivatives of $F(\alpha,\beta,\varphi)$ as we can easily
show. In fact, using the expressions for $\Delta_\alpha^{1/2}$ and
$\Delta_\beta^{1/2}$ in (\ref{DA}) and (\ref{DB}), we have that

\begin{equation}\label{3.71}
\frac{\partial F}{\partial \widetilde E} = -m^2C_2\,, \hskip1cm
\frac{\partial F}{\partial \ell}= C_3\,,\hskip1cm \frac{\partial
F}{\partial E}=t\,,
\end{equation}
as it can be easily checked. Then, the function $F$ can be written
as

\begin{equation}\label{F}
     F\equiv Et-m\widetilde EC_2+\ell C_3\,,
\end{equation}
where $C_2$ and $C_3$ are dependent on $\alpha$, $\beta$ and
$\varphi$, but they are time independent.

\section{Quantum systems}

 From the point of view of quantum mechanics, the Hamiltonian $H$
as well as the integrals of motion $L$ and $\widetilde H$ are
hermitian operators obtained simply by replacing $p_k\to -i\,\p_k$,
$k=1,2,3$, in (\ref{h}), (\ref{l}) and (\ref{hbis}), respectively
(we have taken $\hbar=1$ along this section). Formal hermiticity
follows from the fact that the operators $H$, $L$ and $\widetilde H$
are symmetric in the usual cartesian coordinates. The kinetic parts
of $H$ and $\widetilde H$ are nonsingular quadratic expressions on
positions and momenta and they are self adjoint (\cite{GGNV}). In
addition, we assume that the potentials $U({\bf x})$ and $\Phi({\bf
x})$ satisfy sufficient conditions so that both $H$ and $\widetilde
H$ be self adjoint (as for example that the conditions in the Kato
Rellich theorem be satisfied \cite{KR}).

The conditions (\ref{g}) on the metric  $g({\bf x})$, and
(\ref{uf})-(\ref{potenciales}) on the potential terms $U({\bf
x}),\Phi({\bf x})$ guarantee the commutation relations for these
operators: \be [H,L]=[\widetilde H, L]=[H,\widetilde H]=0\,. \ee We
will look for the simultaneous eigenfunctions $\psi({\bf x})$ of the
three operators

\be (H-E)\psi = (\widetilde H- \widetilde E)\psi = (L - \ell)\psi=0
\ee
or equivalently

\be
\begin{array}{l}
-\Delta \psi({\bf x}) = 2m(E- U({\bf x}))\psi({\bf x})\\[2.ex]
-\widetilde \Delta \psi({\bf x}) =
2m(\widetilde E- \Phi({\bf x}))\psi({\bf x})\\[2.ex]
-i\p_{\varphi} \psi({\bf x}) = \ell\, \psi({\bf x})
\end{array}\,,
\ee
where

\be \Delta = \p_k\p_k, \qquad \widetilde \Delta = \p_j g^{jk}({\bf
x}) \p_k\,. \ee Now, we will express these differential operators in
terms of the coordinates $\l_+,\l_-,\varphi$ in order to rewrite the
eigen-equations in the form

\be\label{sch}
\begin{array}{l}
\displaystyle
-\frac{1}{\l_+-\l_-}\left\{\left[4\l_+(\l_++a^2)\psi_{++} +
2(a^2+3\l_ +)\psi_+ - \frac{a^2}{\l_++a^2}
\psi_{\varphi\varphi}\right]\right.
\\[2.5ex]
\displaystyle \qquad \left.-
\left[4\l_-(\l_-+a^2)\psi_{--}+2(a^2+3\l_-)\psi_{-} -
\frac{a^2}{\l_-+a^2}\psi_{\varphi\varphi}\right]\right\}=
2m(E-U)\psi\\[2.5ex]
\displaystyle -\frac{\l_-}{\l_+-\l_-}\left[4\l_+(\l_++a^2)\psi_{++}
+ 2(a^2+3\l_+)\psi_+ +
\frac{\l_+}{\l_+ +a^2}\psi_{\varphi\varphi}\right] \\[2.5ex]
\displaystyle \qquad
+\frac{\l_+}{\l_+-\l_-}\left[4\l_+(\l_++a^2)\psi_{--} +
2(a^2+3\l_-)\psi_- + \frac{\l_-}{\l_-
+a^2}\psi_{\varphi\varphi}\right]= 2m(\widetilde E- \Phi)\psi
\end{array}
\ee Taking into account that \be \psi_{\varphi\varphi}= -
\ell^2\,\psi \ee and also (\ref{philambda}), equations (\ref{sch})
can be written in a separated form, \be\label{eigeneq}
\begin{array}{l}
\displaystyle -\left[4\l_+(\l_++a^2)\psi_{++} +2(a^2+3\l_+)\psi_+ -
\frac{\ell^2\, \l_+ }{\l_+ +a^2}\, \psi \right] = 2m \left[(E-U)\l_+
- (\widetilde E-\Phi)\right]\psi
\\[2.5ex]
\displaystyle -\left[4\l_-(\l_-+a^2)\psi_{--} +2(a^2+3\l_-)\psi_- -
\frac{\ell^2\,\l_- }{\l_- +a^2}\, \psi \right] = 2m \left[(E-U)\l_-
- (\widetilde E-\Phi)\right]\psi
\end{array}\,,
\ee
where $\lambda_+$ and $\lambda_-$ appear in the first and second
equation in (\ref{eigeneq}) respectively.  Therefore we can look for
a factorized solution for the eigenfunctions as follows:

\be \psi({\bf x}) = \psi^+(\l_+)\psi^-(\l_-)\,e^{i\,\ell\,\varphi}
\ee We can go back to the change of coordinates given by
(\ref{3.47}). This change of coordinates is also suggested by the
formulas below, as we shall see. Now, it is time for choosing
explicit forms for the functions $f(x)$ and $g(x)$ in (\ref{fg}).
For $f(x)$, we shall choose the function that vanish identically.
For $g(x)$, we choose $g(\lambda_-):=-Q(\lambda_-+a^2)$, where $Q$
is a constant. Then, the following expressions arise:

\be
\begin{array}{ll}
\displaystyle f(\l_+) = \Phi-\l_+U = 0,\qquad
&g(\l_-) = \Phi-\l_-U = - Q (\l_-+a^2)\\[2.ex]
\displaystyle U({\bf x}) = -Q\, \frac{\l_-+a^2}{\l_+-\l_-},\qquad &
\displaystyle \Phi({\bf  x}) = - Q\,
\frac{\l_+(\l_-+a^2)}{\l_+-\l_-}
\end{array}\label{4.55}
\ee Or, in cartesian coordinates, \be U({\bf x}) =
-\frac{Q}{2}\,\left( \frac{{\bf x}^2 +a^2}{\left[({\bf x}^2+a^2)^2-
4a^2({\bf x}^2-({{\bf x}\cdot{\bf n}})^2)\right]^{1/2}}-1\right) \ee
Now, we carry  (\ref{3.47}) and (\ref{4.55}) into (\ref{eigeneq}) to
get the following set of two differential equations in the variables
$\alpha$ and $\beta$:

\begin{eqnarray}
  \left\{\frac{d^2}{d \a^2} + \frac{\sinh \a}{\cosh
\a}\frac{d}{d \a} - \frac{\sinh^2 \a}{\cosh^2 \a}\,\ell^2 + 2m
\left[ E\, a^2 \sinh^2 \a - \widetilde E\right]\right\}\psi^+(\a) =0
\,,\label{4.85}
\\[2ex]
  \left\{\frac{d^2}{d \b^2} - \frac{\sin \b}{\cos
\b}\frac{d}{d \b} - \frac{\sin^2 \b}{\cos^2 \b}\,\ell^2 + 2m\left[
E\, a^2 \sin^2 \b + \widetilde E+ Q\, a^2
\cos^2\b\right]\right\}\psi^-(\b) =0\,. \label{4.86}
\end{eqnarray}
This is a special type of equations already studied in the
literature that we briefly analyze in the next section.

%%%%%%%%%%%%%%%%%%%%%%%%%%
\begin{figure}[h]
  \centering
\includegraphics[width=0.4\textwidth]{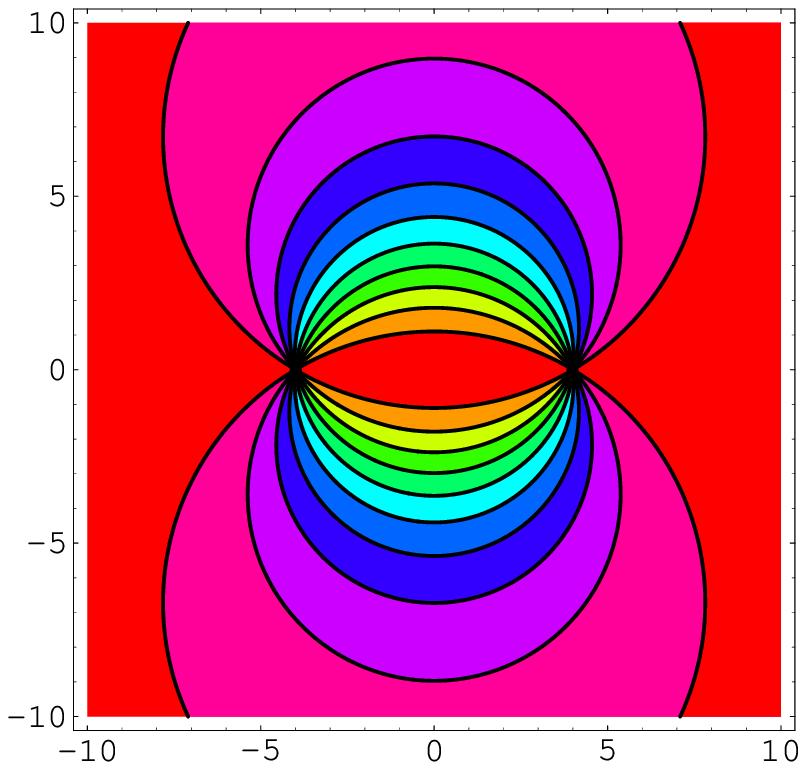}
\hskip0.25cm
\includegraphics[width=0.4\textwidth]{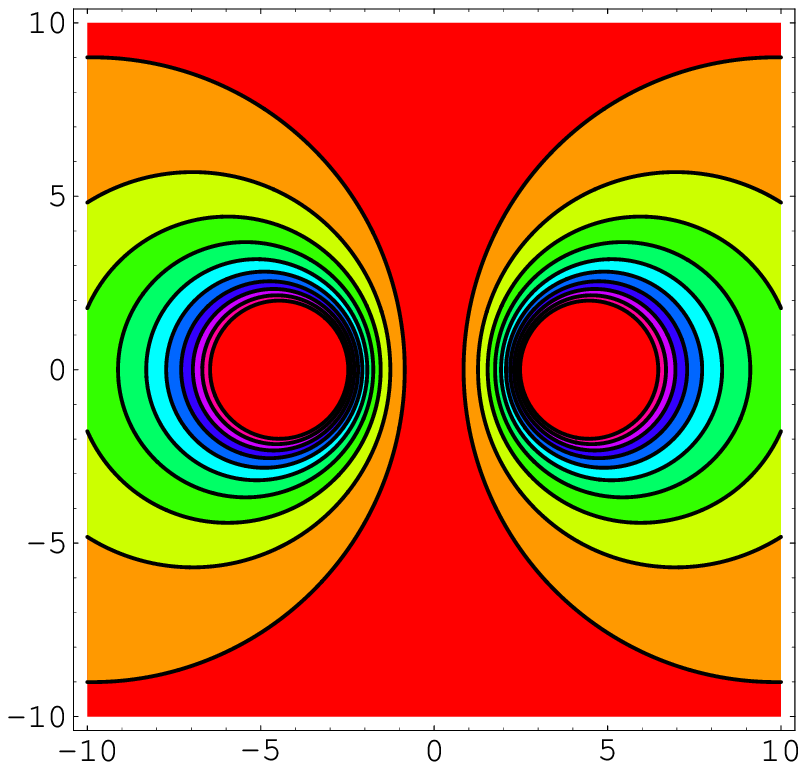}
\hskip0.25cm
\includegraphics[width=0.4\textwidth]{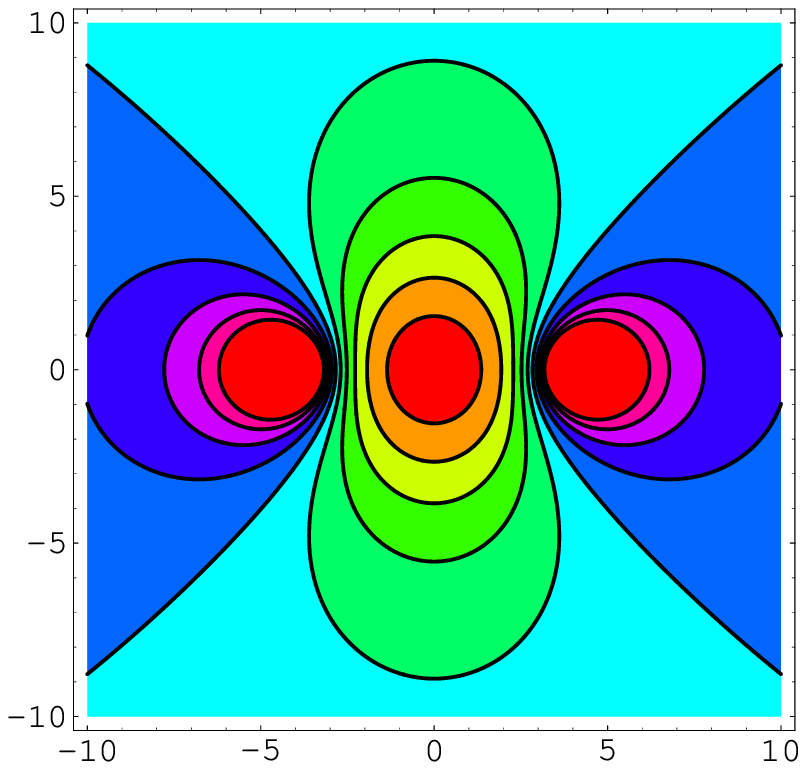}
\caption{Iso-potential lines in the $XZ$-plane  for three values
of the parameters.}
  \label{figura1}
\end{figure}
%%%%%%%%%%%%%%%%%%%%%%%%%%

%%%%%%%%%%%%%%%%%%%%%%%%%%%%%%%%%%%%%%%%%%%%%%%%%%%
\section{Study of the equations and their solutions.}

First of all, it seems convenient to simplify equations
(\ref{4.85}-\ref{4.86}). In order to fulfil this goal, let us choose
the following new coordinates:

\begin{equation}\label{5.87}
    t=\sinh \a,\quad  u=\sin \b\,, \quad \frac{d}{d\,\a}= \sqrt{t^2+1}\,\frac{d}{d\,
    t}\,, \quad \frac{d}{d\, \b}=\sqrt{1-u^2}\,\frac{d}{d\, u}\,.
\end{equation}

Let us introduce the following parameters:

\begin{equation}\label{5.88}
    {\cal E} =2m a^2 E,\qquad q= 2m a^2 Q,\qquad \widetilde{\cal
E} = \ell^2 +2m \widetilde E\,,
\end{equation}
then, eq.\ (\ref{4.85}-\ref{4.86}) become respectively

\begin{eqnarray}
\left[(t^2+1) \frac{d^2}{d t^2} + 2t\,\frac{d}{d t} +
\frac{\ell^2}{t^2+1} +{\cal E} t^2 - \widetilde{\cal
E}\right]\psi^+(t)=0\,. \label{5.89}
 \\[3ex]
 \left[(1-u^2) \frac{d^2}{d u^2} - 2u\,\frac{d}{d u} -
 \frac{\ell^2}{1-u^2} +{\cal E} u^2 +q(1-u^2) +
 \widetilde{\cal E}\right]\psi^-(u)=0\,.
\label{5.90} \end{eqnarray}

We observe that in the case of ${\cal E} = q=0$ and only in this
case, these equations can be reduced to equations of hypergeometric
type, which can be solved in terms of hypergeometric functions.
However, this is not the most general case, let us consider the
following differential equation:

\begin{equation}\label{5.91}
    (1-z^2)\frac{d^2w}{dz^2}-2z\frac{dw}{dz}+\left\{\lambda-\frac{\mu^2}{1-z^2}+\gamma^2
    (1-z^2)\right\}w=0\,,
\end{equation}
where $\lambda$, $\mu$ and $\gamma^2$ are real parameters
($\gamma^2$ may be positive or negative depending on the use of
prolate or oblate coordinates respectively). This is called the {\it
spheroidal wave equation} \cite{fma}. It is very simple to show that
both equations (\ref{5.89}) and (\ref{5.90}) are versions of the
spheroidal wave equation. In fact, (\ref{5.90}) can be written as

\begin{equation}\label{5.92}
(1-u^2) \frac{d^2\psi^-(u)}{d u^2} - 2u\,\frac{d\psi^-(u)}{d u}
+\left\{ G- \frac{\ell^2}{1-u^2}  +q'(1-u^2)\, \right\}
\psi^-(u)=0\,,
\end{equation}
with $G:=\widetilde{\cal E}+{\cal E}$ and $q'= q-{\cal E}$. If we
apply the change of variables given by $t=i\alpha$, equation
(\ref{5.89}) becomes

\begin{equation}\label{5.93}
(1-\alpha^2) \frac{d^2\psi^+(\alpha)}{d \alpha^2} -
2\alpha\,\frac{d\psi^+(\alpha)}{d \alpha}
+\left\{G-\frac{\ell^2}{1-\alpha^2}-{\cal
E}(1-\alpha^2)\,\right\}\psi^+(\alpha)\,,
\end{equation}
where again $G=\widetilde{\cal E}+{\cal E}$ and we have kept the
notation $\psi^+(\alpha)=\psi^+(i\alpha)=\psi^+(t)$.

Solutions of the spheroidal wave equation (\ref{5.91}) and therefore
of (\ref{5.92}) and (\ref{5.93}) have been studied in \cite{fma}.
The origin $z=0$ is a regular point of the equation and therefore,
we can find two linearly independent functions in terms of powers
series on the variable $z$. These series converges on the open
circle centered at the origin and radius equal to one, since $z=\pm
1$ are singular points for the equation. On this open circle, one
can find one even and one odd solution of (\ref{5.91}) of the form
$\sum_{n=0}^\infty a_{n}z^{2n}$ and $\sum_{n=0}^\infty a_nz^{2n+1}$
respectively, which are linearly independent. These series do not
converge at the singular points $\pm 1$. As they do converge on the
open interval $(-1,1)$, the wave function $\psi^-(\beta)$ solution
of equation (\ref{4.86}) is periodic on the real axis with
singularities at the points $(2n+1)\pi/2$. There exists another type
of linearly independent even and odd solutions on a neighborhood of
the origin that may converge at the singular points $\pm 1$,
provided that a relation is satisfied between the coefficients
$\lambda$, $\mu$ and $\gamma$ in (\ref{5.91}) (and its corresponding
translation in terms of the coefficients in (\ref{5.92}) and
(\ref{5.93})) \cite{fma}. In any case, the solutions $\psi^-(\beta)$
of (\ref{4.86}) on the real axis are periodic and therefore, not
square integrable.

With respect to equation (\ref{5.91}), $z=\pm 1$ are regular
singular points with indices equal to $\pm\frac 12\mu=\pm\frac
12\ell$. Being $\ell$ an integer, the two linearly independent
solutions on a neighborhood of $z=1$ are $u_1(z)$ and $u_2(z)$ with
$u_1(z)=(z-1)^{\ell/2}F_1(z-1)$ and
$u_2(z)=u_1(z)\log(z-1)+(z-1)^{-\ell/2}F_2(z-1)$, where $F_1(z-1)$
and $F_2(z-1)$ are power series on $z-1$. These power series have
radii of convergence equal to 2. On a neighborhood of $z=-1$,
similar solutions can be found. Power series never truncate.

There is another singular point at $z=\infty$. This singular point
is irregular. Solutions on a neighborhood of the infinite have the
form $z^\nu \sum_{n=-\infty}^\infty a_nz^{2n}$, where $\nu$ is a
complex number depending on the equation parameters. In order to
simplify the recurrence relations for the coefficients $a_n$, it is
customary to choose this solution as
$(z^2-1)^{\mu/2}z^{\nu-\mu}\sum_{n=-\infty}^\infty a_nz^{2n}$\,. The
condition that the Laurent series converges in $1<|z|<\infty$ gives
a relation between $\lambda,\mu,\gamma$ and $\nu$ \cite{fma}. These
solutions are of the form

\begin{equation}\label{5.94}
(z^2-1)^{\mu/2}\, z^{\mu}\sum_{n=-\infty}^\infty a^\mu_{\nu,n}
\psi^{(j)}_{\nu+2n}(\gamma z)\,,\qquad j=1,2,3,4\,,
\end{equation}
where

\begin{equation}\label{5.95}
\psi^{(j)}_\nu(z)=\left(
\frac{\pi}{2z}\right)^{1/2}\;Z^{(j)}_\nu(z)\,,
\end{equation}
with $Z^{(1)}_\nu(z)=J_\nu(z)$, $Z^{(2)}_\nu(z)=Y_\nu(z)$,
$Z^{(3)}_\nu(z)=H^{(1)}_\nu(z)$ and $Z^{(4)}_\nu(z)=
H^{(2)}_\nu(z)$, being $J_\nu(z),Y_\nu(z)$ and $H^{(i)}_\nu(z)$ the
Bessel functions of first, second and third class respectively. Any
two of the set of solutions (\ref{5.94}) are linearly independent
provided that $\nu$ be not a half odd integer.

Solutions of (\ref{5.89}) and (\ref{5.90}) can be obtained without
resorting to the standard study of the spheroidal wave function. For
instance, if we use the change of variables given by $z:=t^2$ in
(\ref{5.89}), this equation is transformed into

\begin{equation}\label{5.96}
   \left[ 4z(z+1)\,\frac{d^2}{dz^2}+(6z+2)\,\frac{d}{dz}
    +\frac{\ell^2}{z+1}+{\cal E}z-\widetilde{\cal E}
    \right]\psi(z)=0\,.
\end{equation}
Now, the singular regular points lie at $z=0,-1$ and it is not
difficult to obtain solutions in form of power series on a
neighborhood of these points. For example, for $z=0$ the
characteristic exponents are $0$ and $1/2$ giving respective
linearly independent solutions of (\ref{5.96}) of the form
$\psi_0(z)=\sum_{n=0}^\infty a_n z^n$ and
$\psi_{1/2}(z)=\sum_{n=0}^\infty b_nz^{n+\frac 12}$. Recurrence
relations for the coefficients depend on four coefficients, except
the first and second relations which depend on the two and three
first coefficients respectively (which is compatible with the fact
that $a_0$ and $b_0$ should be the only independent coefficients).
On a neighborhood of the singular point $z=-1$, two linearly
independent solutions can be found of the form
$\psi_1(z)=\sum_{n=0}^\infty a_n(z+1)^{n+\ell/2}$ and
$\psi_2(z)=\psi_1(z)\,\log(z+1)+\sum_{n=0}^\infty
 b_n(z+1)^{n-\ell/2}$. These series make sense provided that
compatibility relations exists between the parameters $\ell$, ${\cal
E}$ and $\widetilde{\cal E}$ in complete agreement with the general
study of the solutions of spheroidal wave functions in \cite{fma}.

\section{Conclusions and remarks.}

We have studied the conditions of integrability of a classical or
quantum system having a symmetry axis. As in a three dimensional
integrable system, we have found three independent observables such
that their respective Poisson brackets are zero, in the classical
case, or commute in the quantum case. The chosen symmetry forces one
of the observables to be the component of the angular momentum in
the direction of the symmetry axis. The other two can be written in
Hamiltonian form as a sum of a kinetic term plus a potential.

In the classical case, we have obtained the most general form of the
potentials corresponding to both Hamiltonians in terms of oblate
spheroidal coordinates, that depends on two arbitrary functions
depending on one coordinate only.  We have written the equations of
motion in terms of this coordinates and show that the
Hamilton-Jacobi characteristic function can be written as a sum of
three functions each one depending on one coordinate only. Then, we
have obtained the explicit form for these three functions.

The quantum case is obtained by direct canonical quantization of the
classical case. The condition of integrability yields to two
Schr\"odiger type equations in which with separate variables. Then,
a reasonable choice on the functions that determine the potentials
yields to new equations that are shown to be of the spheroidal type.
We finish the discussion with some comments on the solutions of this
kind of equations.

%%%%%%%%%%%%%%%%%%%%%%%%%%%%%%%%%%%%%%%%%%%%%%%%%
\section*{Acknowledgments}

We are grateful to Profs. M. Ioffe,  L.P. Lara and M. Santander for
useful comments. Partial financial support is acknowledged to the
Junta de Castilla y Le\'on Project VA013C05, the Ministry of
Education and Science of Spain projects MTM2005-09183 and
FIS2005-03988 and Grant SAB2004-0169 and the Russian Science
Foundation Grant 04-01-00352.

%%%%%%%%%%%%%%%%%%%%%% BIBLIOGRAPHY %%%%%%%%%%%%%%%%%%%%%%%%%%%

\end{document}